# Self-organized growth of Ag islands on Si(111)-(7×7) – optimization of an STM experiment by means of KMC simulations

Pavel Kocán[1], Pavel Sobotík[1], Ivan Ošťádal[1], Miroslav Kotrla[2]

[1]Department of Electronics and Vacuum Physics, Charles University, V Holešovičkách 2,
180 00 Prague 8, Czech Republic
sobotik@mbox.troja.mff.cuni.cz
[2]Institute of Physics, Academy of Sciences of the Czech Republic, Na Slovance 2, 182 21 Praha 8, Czech Republic

**Abstract**

A growth model and parameters obtained in our previous experimental (STM) and theoretical (Kinetic Monte Carlo simulations) studies of Ag/ Si(111)-(7×7) heteroepitaxy were used to optimise growth conditions (temperature and deposition rate) for the most ordered self-organized growth of Ag island arrays on the (7×7) reconstructed surface. The conditions were estimated by means of KMC simulations using the preference in occupation of half unit cells (HUCs) of F-type as a criterion of island ordering. Morphology of experimentally prepared island structures was studied by STM. High degree of experimentally obtained island ordering is compared with the simulated data and results are discussed with respect to the model and parameters used at the KMC simulations.

**Keywords**

heteroepitaxy, silver, Si(111)-(7×7), KMC, scanning tunneling microscope, self-ordering

**Introduction**
A lot of effort has been paid to prepare ordered arrays of metal nanoclusters because of interesting possible applications in the field of microelectronics. Reconstructed Si(111)-(7×7) surface was used as a template for spontaneous ordering. Recently ordered arrays of In, Al, Ga identical clusters have been prepared [1,2]. Highly ordered arrays of identical bimetal clusters of In/Ag and In/Mn were also reported in [2]. Growth of the identical clusters is mediated by the existence of particular magic sizes of atom clusters that are more stable than the others. In case of Tl [3], Sn [4] and Ag [5,6] more or less ordered island arrays were reported but magic clustering was not observed. Preparation of ordered structures requires a delicate control of deposition conditions. Kinetics of self-ordering of clusters on the reconstructed surface depends on many parameters and theoretical description of a role of deposition conditions is difficult and still not available. Kinetic Monte Carlo simulation (KMC) of the growth represents an effective approach for theoretical study of important processes involved. We used the KMC simulations together with growth experiments for studying nucleation and heteroepitaxial growth of Ag on the Si(111)-(7×7) surface [5,6,7,8].

The (7×7) reconstructed surface consists of triangular half-unit cells (HUCs), which represent potential wells for adsorbed atoms hopping on the surface [9,10]. Each HUC contains six Si adatoms of a top layer. The HUCs are of two types, one containing structural fault (FHUC), the other is unfaulted (UHUC) [11]. It results in different reactivities and consequently in preferential nucleation of Ag in FHUCs. The Ag adatom hopping between various adsorption positions within a HUC at room and higher temperatures is easy but the mobility of the Ag adatoms on the surface – important for growth processes – is determined by the hopping rate between HUCs (depends on barrier height). In a coarse grained model, we used for the simulations, the surface is represented by adsorption sites of two types (FHUCs, UHUCs) with different barrier heights. The preference in FHUC occupancy, $P_F$, (ratio of number of FHUCs occupied by Ag clusters to number of all occupied HUCs) is determined by the barrier difference.

Metal clusters on Si(111)-(7×7) surface can be ordered in two ways – they can occupy either HUCs of both types (honey comb symmetry) or HUCs of one type only (triangular symmetry). The second case reflects the preferential adsorption of metal adatoms in one type of HUCs and the preference $P_F$ is a measure of cluster ordering.

In this paper we focus on finding optimum deposition conditions for the most ordered arrays of Ag clusters on the Si(111)-(7×7) surface. We use the growth model developed in our previous studies and parameters obtained from KMC simulations [6,8,12]. We performed computer simulations to study the influence of deposition parameters (temperature and deposition rate) on island ordering and visualize it. The series of samples was prepared at various temperatures to verify simulated dependencies. The film morphology was studied by STM.

**KMC model**
The coarse grained KMC model implements the following growth scenario:
(i) Random surface positions (HUCs) are selected for atoms impinging with rate $F$.
To mimic the short-range ordering of Ag-object and the low value of coverage (see [5,6] for details) at RT, when thermally activated hopping is negligible, the transient mobility is implemented according to [6] as follows:
Atoms deposited into a HUC are immediately moved to adjacent occupied HUC, if it exists, otherwise they stay with 100 % probability in F HUC and 95 % probability in U HUC (to reproduce preference at RT) [1].

(ii) Single adsorbed Ag atoms (Ag-adatoms) jump between adjacent HUCs and create nuclei (clusters).

(iii) When number of atoms in a cluster $n \geq n^*$ ($n^*$ - critical size, the same for F and UHUCs) the cluster can decay. The activation energy for hop out of HUC containing $n$ atoms is given by $E_n = E_{F,U} + (n - 1)E_a$ for $n \geq n^*$ ($E_{F,U}$ is the barrier for single Ag atom hop out of FHUC or UHUC respectively, $E_a$ is the effective binding energy between adatoms inside a HUC). Rate of the corresponding process is $\nu_n^{F,U} = n\nu_0^{F,U}\exp(-E_n/kT)$ ($\nu_0^{F,U}$ is the frequency prefactor, $k$ Boltzmann constant and $T$ sample temperature).

(iv) The capacity of a HUC to accommodate atoms is limited; clusters containing $n_S$ atoms do not capture further adatoms – isolated cluster cannot overgrow HUC boundaries (see [7,8]).

(v) Further model features, which involve processes at formation of larger islands (during the growth up to 0.6 ML), are described in details in [7,8], but they are not important for the structures studied in this paper.

The model parameters used for simulations: $\nu_0^F = \nu_0^U = 5 \times 10^9$ s$^{-1}$, $E_F = 0.70$ eV, $E_U = 0.67$ eV, $E_a = 0.05$ eV, $n^* = 5$, $n_S = 21$. Details of the model implementation were described in [6], simulations of larger islands growth will be presented elsewhere [8].

**Experimental details**

We used an UHV STM of our design and construction (base pressure $< 5 \times 10^{-9}$ Pa) for in-situ measurements. Samples with thickness $(0.11 \pm 0.03)$ ML ($\approx 2.5$ atoms per HUC, 1 ML $= 7.83 \times 10^{14}$ atoms/cm$^2$) were prepared at substrate temperatures from 340 K to 390 K (absolute calibration error $\pm 10$ K) and at deposition rate $R = 0.0006$ ML/s. Ag was evaporated from the tungsten wire. The deposition rate was measured by a quartz thickness monitor (absolute accuracy $\pm 10\%$). Clean Si(111)-(7×7) surface was obtained using standard flashing procedure. Substrates (Sb doped, resistivity $0.005 \div 0.01$ $\Omega$cm) were heated by passing DC current. Samples relaxed before STM measurements at least one hour at room temperature (RT).

**Results and discussion**

The Ag adatoms preferentially occupy FHUCs, which is energetically the most favourable (for low amounts of deposited material). The preference $P_F$ is kinetically determined and depends on a ratio between hopping rate and deposition flux. Maximum value of $P_F$ can be achieved when Ag adatoms have enough time for rearranging on the surface during the growth. Such conditions are considerably supported by presence of unstable Ag clusters – i.e. when amount of deposited Ag corresponds to $n^*$ Ag atoms per a couple of HUCs ($\approx 0.1$ ML) at the most.

The simulated dependences of the preference $P_F$ and the coverage (ratio of occupied to the total number of HUCs on the surface) on substrate temperature for various deposition rates are in Fig. 1. For every deposition rate a temperature $T_M$ corresponding to a maximum of the preference $P_{FM}$ can be found. At temperatures $T < T_M$ lower mobility of Ag adatoms results in lowering of opportunity to occupy FHUCs. Increasing mobility at temperatures $T > T_M$ at given deposition rate results in growth of larger islands covering two or more HUCs, which decreases the preference. Lowering the maxima $P_{FM}$ with higher values of $T_M$ in Fig.1 corresponds to decreasing role of the barrier difference between FHUCs and UHUCs with increasing temperature [14].

---

[1] Due to the unclear origin of the transient mobility mechanism a correlated diffusion [13] or biased diffusion was offered as an alternative mechanism [5,6], but this mechanism is still not able to explain simultaneously observed Ag morphological features [5,6] (especially ordered F and U domains) together with observed dynamical behaviour of single Ag adatoms [14]. Final decision requires a new set of carefully designed experiments to accept or abandon the idea of correlated diffusion

The temperature $T_M$ increases with the deposition rate – Fig. 2a. The almost linear dependence in a semi-logarithmic scale shows that the ratio between the deposition and hopping rates is constant for the best ordering conditions. Decreasing of $P_{FM}$ with the rate (Fig. 2b) means that the deposition should be as slow as possible at an experiment – the temperature corresponding to the deposition rate is given by the diagram in Fig. 2a.

STM images of films prepared at the theoretically obtained conditions show high degree of cluster ordering – Fig. 3a, b. Morphology of samples prepared at temperatures higher than $T_M$ is in Fig. c, d. Experimental dependences of the preference and coverage on the substrate temperature are compared with the simulated data in Fig. 4a, b. An agreement between the coverage dependences is very good. The temperature $T_M$ obtained from the experiment differs from the simulated value within a range of error estimated for temperature measurement. The experimental value of $P_{FM}$ is lower than the theoretical one. The dependences on temperature quantitatively differ also. We can understand the difference as a measure of proximity of the model and real processes under conditions when the sensitivity of the growth on conditions is extreme. The simulations indicated considerable dependence of the preference on a number of atoms in unstable cluster – $n^*$.

STM images cannot provide numbers of Ag atoms in clusters, only monomers and dimers can be distinguished [9]. Cluster size distributions can be obtained from simulations. The simulations showed that the investigated structures are unstable even after several hours of relaxation at RT (there are still monomers and dimers present on the surface). It was confirmed experimentally from STM images as well. Stability of the ordered structures can be increased by annealing (few minutes at 500K [6] ) which results in decay of unstable clusters and attachment of hopping adatoms to the stable ones. Further deposition stabilize the structure as well – existing clusters capture deposited Ag adatoms until the saturated value $n_S$ is reached.

**Conclusions**

The growth conditions for the most ordered films were found by means of the KMC model and verified by the STM measurements. The ability of the model to show the influence of deposition conditions on the film morphology was demonstrated.

The model can be used for other non-reactive metals after adequate adjustment of the model parameters and, if necessary, definition of more complex dependence of the activation energy $E_n$ on number of atoms in a HUC. Though we did not find any experimental support for magic clustering at Ag/Si(111)-(7×7) heteroepitaxy at room and higher temperatures, such mechanism can be implemented into the model via $E_n(n)$.

**Acknowledgments**

The presented work was supported by the Grant Agency of Czech Republic - projects 202/01/0928 and 202/03/0792, and by the Ministry of Education, Youth and Sports of Czech Republic - project FRVŠ 2735/2003.

**Figures**

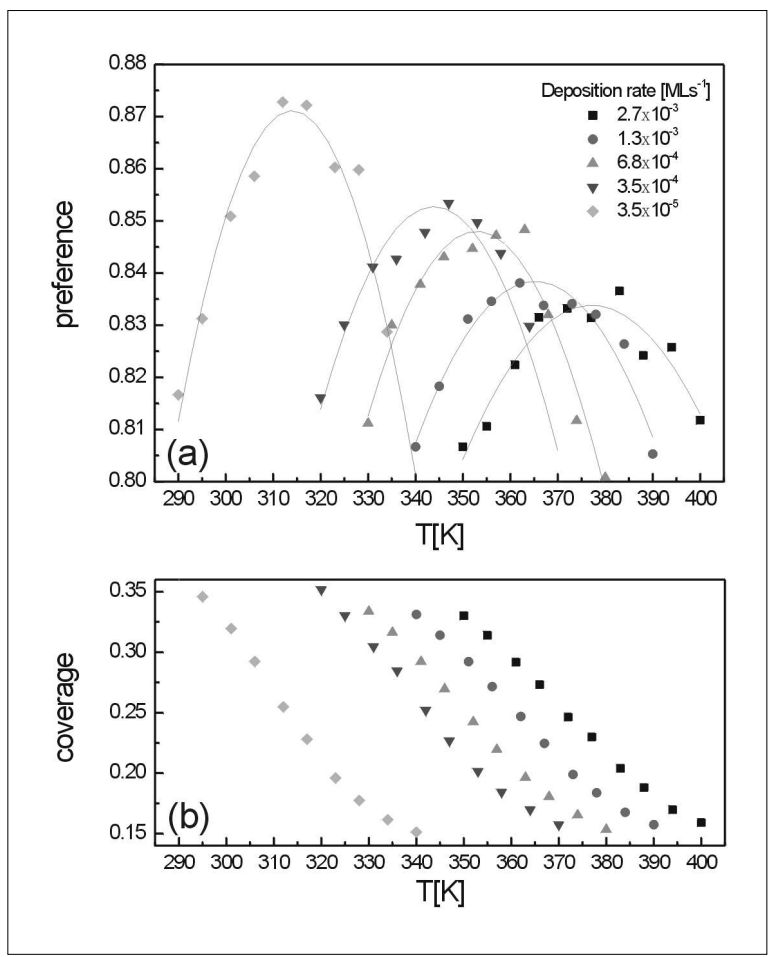

Fig. 1 – Simulated dependences of preference in FHUC occupation (a) and HUC coverage (b) on deposition temperature $T$ for films prepared at various deposition rates. Polynomial fitting (lines) was used to highlight maxima in the preference at the panel (a). Deposited amount $d = (0.11 \pm 0.03)$ ML corresponds to 2.5 atoms per HUC.

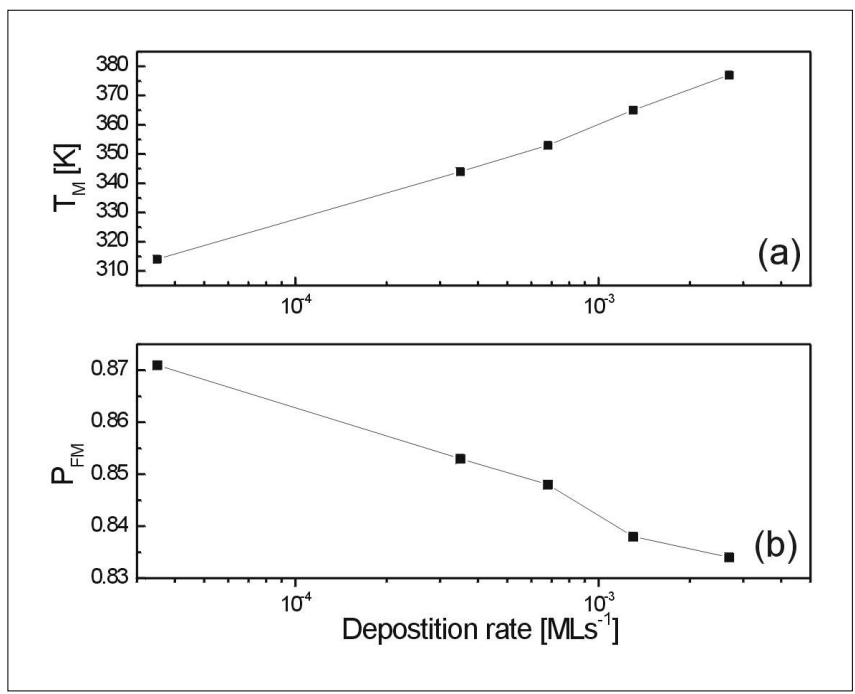

Fig. 2 – Temperature $T_M$ (a) and preference $P_{FM}$ (b) corresponding to the most ordered films as functions of deposition rate. Conditions are the same as in Fig. 1.

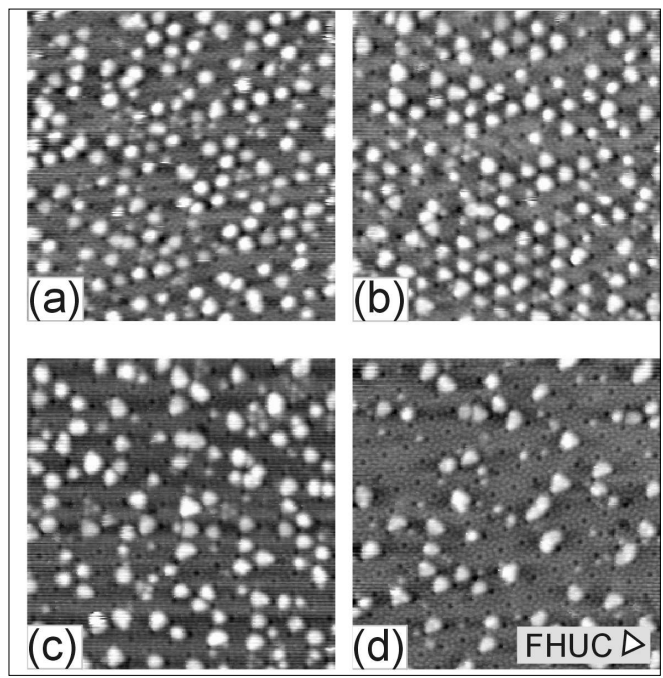

Fig. 3 – STM images (35×35 nm$^2$, tip voltage -2 V, tunneling current 0.4 nA) of films deposited at the rate $F$ = 0.0006MLs$^{-1}$. Deposited amount $d$ = 0.11 ML, substrate temperatures: (a) $T$ = 346 K, (b) $T$ = 354 K, (c) $T$ = 372 K, (d) $T$ = 384 K. Orientation of FHUCs (the same for all frames) is marked in the frame (d). Morphologies (a) and (b) show high degree of cluster ordering.

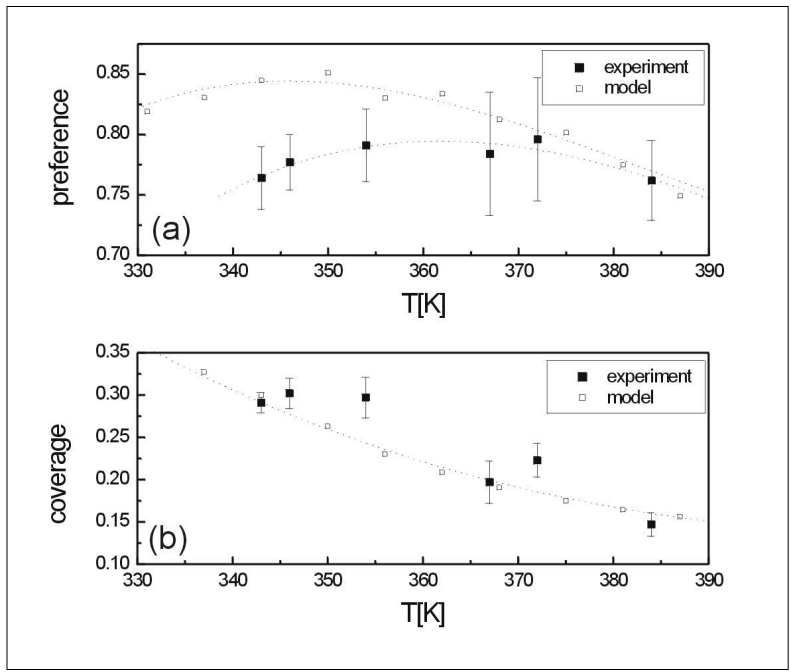

Fig. 4 – Comparison of experimental and simulated data for films ($d = 0.11$ ML) deposited at the rate $F = 0.0006 \text{MLs}^{-1}$. (a) – preference in FHUC occupation, (b) – HUC coverage as functions of the substrate temperature.